# The Sleeping Beauty Problem – A Real-World Solution

Hutan Ashrafian

*Institute of Global Health Innovation, Imperial College London, Leeds University Business School and Institute of Civilisation*

**Correspondence to:**

Professor Hutan Ashrafian

Institute of Global Health Innovation, Imperial College London,

10th Floor Queen Elizabeth the Queen Mother (QEQM) Building,

St Mary's Hospital,

Praed Street, London,

W2 1NY, UK

h.ashrafian@imperial.ac.uk; hutan@ic.ac.uk

**Abstract**

The Sleeping Beauty Problem remains a paradoxical problem that penetrates multiple disciplines that include probability theory, self-locating belief, decision theory, cognitive science, the philosophy of mathematics and science. It asks Sleeping Beauty's credence of a coin toss being heads in the experiment that incites two main stances, that of the 'Halfers' and 'Thirders'. Here a real-world empirical approach numerically highlights breakdown between these groups and considers the role of how a real-world application of such an experiment with sleep induction by anesthesia and pharmacological amnesia induction would affect the probability of Sleeping Beauty's credence.

**Introduction**

The Sleeping Beauty Problem remains a controversial issue in the fields of probability theory, self-locating belief, decision theory, cognitive science, the philosophy of mathematics and science. It presents a seemingly uncomplex scenario[1] with a direct question of probability set out below.

An ethically compliant experiment is executed where a random dichotomous variable outcome is achieved with a fair coin toss to achieve Heads or Tails. A young female volunteer who is a rational epistemic agent (known as Sleeping Beauty) is sleep induced on a Sunday with the following study protocol:

(i)[Heads – one interview] If the coin turns up heads, then Sleeping Beauty is sleep induced from the Sunday to Monday then woken up and interviewed without being given access or context to her waking time. She is then sleep induced from Monday to Wednesday.

(ii)[Tails – two interviews] If the coin turns up Tails, then Sleeping Beauty is sleep induced from the Sunday to Monday then woken up and interviewed without being given access or context to her waking time. She is then sleep induced from Monday to Tuesday then woken up and interviewed without being given access or context to her waking time, and once again sleep induced from Tuesday to Wednesday.

(iii)[Final waking and interview on Wednesday] Whether having undergone either path (i) or (ii), she is woken up on Wednesday, again without being given access or context to her waking time and undergoes one final interview.

(iv)The interview question on all interview occasions is he question: What is Sleeping Beauty's credence (degree of belief) that the coin toss at experiment initiation was heads?

There are two main stances aiming to offer a solution to the question of Sleeping Beauty's belief of whether the coin was heads at the end of the experiment [2]–[4]. The 'Halfers' position is that as the coin is fair, independent to the number of times that sleeping beauty is woken (which she herself won't know), the underlying probability of the coins dichotomy will prevail with a probability of Heads as one half. Conversely, the 'Thirders' position is based on Bayesian inference and/or the fact that there is one interview for heads versus two interviews for Tails before experiment end, then the chance of heads would be one third. Both these positions remain entrenched, even with the application of extensions to the thought experiment such as the Extreme Sleeping Beauty experiment where is she awakened a million times [5], the double sleeping beauty problem and the infinite double sleeping problem [6]. The aim of this investigation therefore was help address these multiple issues by a practical real-world solution approach of what would Sleeping Beauty's credence be, if she was to actually undergo such a test.

**Methods – A Real World Approach**

Sleep Methodology through Anesthetic Agents

It has typically been surmised that Sleeping Beauty would be sleep induced by a sleep induction agent, typically by 'sleeping pills'. Each time she sleeps, it will be for 24 hours before being woken. In order to achieve that level of persistent sleep for that period of specific time safely, clinically induced sleep will be necessary. Currently sleeping pills administered in this manner aren't adequate for this purpose and she will require general anesthesia (rather than local, regional, or sedative anesthesia), most likely with an intravenous (IV) administered agent. For waking at the pre-specified time, subsequent pharmacological anesthesia would be required for the proposed sleep cycles.

Sleep is a complex multimodal set of physiological brain-initiated events converging toward a state of non-wakefulness [7], [8]. It is run by (i) a timing cycle through a 24h circadian rhythm system and (ii) a homeostatic feedback mechanism to accommodate depth and duration in response to need. Together these lead to a dichotomous split between REM (rapid eye-movement) and NREM (non-rapid eye-movement) sleep periods. REM is associated with the cholinergic system drive, whereas NREM is associated with GABAergic/galanin activity in the region of the ventrolateral preoptic nucleus (VLPO).

Anesthetic induced sleep is biochemically distinct from 'natural sleep' and utilizes some similar though distinct pathways [7], [8]. For example, those of GABA-mediated channel (such as $GABA_A$) activation (propofol, barbiturates, etomidate), those inhibiting N-methyl-D-aspartate (NMDA) receptors (Ketamine, $N_2O$, Xenon), and central $α_2$ adrenergic receptors (Dexmedetomidine). General anesthesia typically induces NREM predominant sleep (so that subjects have a subsequent REM rebound) and the depth of unconsciousness (NREM stages 3 and 4) is deeper than 'natural sleep'.

Effects of Sleep Induction and Reversal on Sleeping Beauty's Cognition

Prolonged general anesthesia can result in notable biochemical and neurophysiological sequelae such as recognized post-anesthetic delirium (complicated by the issue that most anesthesia is associated with surgery performed on a pathology), so that the delirium can derive from the effects of anesthesia in the context of surgery and pathology. This renders it difficult to appraise the exact proportion of delirium for anesthesia alone, though nonetheless carries a plausible independent neurocognitive impact. In non-cardiac surgery (generally less physiological impact than the profound effects of cardiac intervention and associated cardiac bypass), the post-operative/anesthetic delirium probability is 18-20% [9], [10]. The chance of this occurring is much higher in cases of prolonged and repeat anesthesia as in the case of sleeping beauty, and the probability of post-operative/anesthetic delirium increases with age [10]. Here, the process of general anesthesia could also induce antegrade amnesia (what goes on after the anesthesia) and retrograde amnesia (what went on before anesthesia).

Delirium is a syndrome of acute brain dysfunction characterized by inattention and other mental status impairments [9]. Added with amnesia, this would impair Sleeping Beauty's capacity to calculate the probability of the coin being Heads or Tails in the question. The effects of drug-induced anesthetic delirium would have consequences of probability discounting capability, including higher cognitive analysis, so that anesthetic associated cognitive dysfunction may carry a higher impact on 'more complex' concepts such as Bayesian inference and differential probabilities [11], [12], possibly including the 'Thirder' concept for the problem.

Probability Calculation

Data for probability of choice in the Sleeping Beauty Problem derives from a sample numerical rating-scale based questionnaire with outcome results and dispersion used for imputing a breakdown of answers to the problem [13]. Here there was groupings for 'Heads', 'Tails' and 'Don't Know'. Subsequent calculation of probabilities to incorporate sleep induction and reversal required in the problem derive from the assumption that the delirium associated with sleep induction and reversal by the physiological effects of anesthesia will likely lead to confusion. Therefore, the likelihood of choosing Heads or Tails will be decreased by the likelihood of delirium, mathematically this would be subtracted equally from the Heads and Tails group and added to the 'Don't Know' group. Additionally a 'forced' dichotomization into Heads and Tails was made for contextual comparison with previous studies on the Sleeping Beauty Problem regarding proportions of respondents answers into this dichotomy.

**Results**

Applying the numbers for *N* from real-world data we have:

(i) In non-anesthetised beauty

Heads: $\frac{2}{6} \approx 33.3\%$
Tails: $\frac{1}{6} \approx 16.6\%$
Don't know: $\frac{3}{6} = 50\%$

Forced dichotomisation:
Heads= $\frac{7}{12} \approx 58.3\%$
Tails= $\frac{5}{12} \approx 41.6\%$

(ii) In anesthetised beauty aged under 65 year (upto 6% risk of delirium)

Heads: $\frac{2}{6} - \frac{3}{100} = \frac{91}{300} \approx 30\%$
Tails: $\frac{1}{6} - \frac{3}{100} = \frac{41}{300} \approx 13.6\%$
Don't know: $\frac{3}{6} + \frac{6}{100} = \frac{14}{25} = 56\%$

Forced dichotomisation:
Heads= $\frac{7}{12} \approx 58.3\%$
Tails= $\frac{5}{12} \approx 41.6\%$

(iii) In anesthetised beauty aged 65-85 years (16% risk of delirium)

Heads: $\frac{2}{6} - \frac{20}{100} = \frac{17}{75} \approx 22.6\%$
Tails: $\frac{1}{6} - \frac{20}{100} = \frac{17}{150} \approx 11.3\%$
Don't know: $\frac{3}{6} + \frac{16}{100} = \frac{66}{100} = 66\%$

Forced dichotomisation:
Heads= $\frac{167}{300} \approx 55.6\%$
Tails= $\frac{133}{300} \approx 44.3\%$

<u>(iv) In anesthetised beauty aged over 85 (40% risk of delirium)</u>

Heads: $\frac{2}{6} \cdot \frac{20}{100} = \frac{1}{15} \approx 6.6\%$
Tails: $\frac{1}{6} \cdot \frac{20}{100} = \frac{1}{30} \approx 3.3\%$
Don't know: $\frac{3}{6} + \frac{40}{100} = \frac{90}{100} = 90\%$

Forced dichotomisation:
Heads = $\frac{31}{60} \approx 51.6\%$
Tails = $\frac{29}{60} \approx 48.3\%$

**Discussion**

A real-world approach to the Sleeping Beauty Problem reveals some interpretable results for this paradox. It also presents an opportunity to dissect out some of the uncertainties encountered in this problem that also serve to offer a springboard for further advances to achieving solutions for this persisting puzzle.

Here practicality presented (i)advances beyond the 'Halfer' and 'Thirder' positions, where both stances are balanced by a 'Don't Know' status. Furthermore (ii) when 'Halfer' and 'Thirder' stances are considered in comparison, the 'Halfer' answer is more likely by approximately a ratio of 2:1 that persists through all age groups in a theoretical Sleeping Beauty.

The concept at the centre of the Sleeping Beauty Problem, namely induced sleeping and pre-determined waking specifically take place through (iii) anesthetic sleep induction which in turn has biological consequences. This can disrupt the nature of beauty's opinion regarding whether the coin toss was heads or tails. A consequence of this pharmacological anesthesia is that the older the beauty is, the higher the likelihood of having post-sleep and waking delirium and therefore confusion.

As a result, if Sleeping Beauty is over 85 years old, she wouldn't know whether the coin was Head or Tails in 90% of cases. (iv) Forced dichotomisation suggests that if we are to statistically separate Sleeping Beauty's opinion into only Heads or Tails, a younger Sleeping Beauty (under 65 years old) would favour an approximate 3:2 Heads favouring Tails, though as Sleeping Beauty ages beyond 85, this asymptotes toward a 1:1 ratio of Heads to Tails opinion.

An approach of forced dichotomization does however have many statistical and methodological drawbacks as it detracts from the original data dispersion and can lead to overtly biased interpretations devoid of its original data collection. In this analysis, it was notable to reveal that at the theoretical results of Sleeping Beauty without anesthesia and those with 6% confusion based on anesthesia offered the same result of forced dichotomisation of heads ($\frac{7}{12}$) versus tails ($\frac{5}{12}$).

As an interesting aside, there is an established medical sleep disorder known as Sleeping Beauty Syndrome [14], [15] or Kleine-Levin Syndrome (KLS) where the is excessive sleep (hypersomnolence), eating disorder (hyperphagia of anorexia), disinhibited behavior (such as hypersexuality), emotional lability, altered perception (derealization) and cognitive dysfunction, which has individuals present some pathologically induced parallels to the probabilistic Sleeping Beauty problem presented herein.

**Conclusion**

A real-world approach to the Sleeping Beauty problem offers some novel insights into overcoming the ambiguity of its presentation to achieve tangible figures of outcome. Here the issue of 'Halfers' and 'Thirders' on their opinion of whether Sleeping Beauty has a credence that the coin is tossed heads needs to be contextualized within the status that "don't know" also exists. Controlling Sleeping Beauty's sleep and wakening based on a coin toss does have the biological consequences of pharmacological anesthesia and reversal with its effects on confusion in decision-making. Here an older Sleeping Beauty will more likely tend toward "I don't know" to the question of the paradox, though if suggesting heads or tails, the choice of heads is more likely across all ages for Sleeping Beauty. Further real-word data through wider data collection of surveys and questionnaires, simulations and avante garde data linkage and artificially intelligent analytical approaches may yet offer even more accurate solutions to this problem.

Compliance with ethical standards

Financial disclosure: The author has no financial conflicts or interests to report in association with the contents of this paper.

Conflict of interest: HA is Chief Scientific Officer, Preemptive Medicine and Health, Flagship Pioneering.

Ethical approval: This article does not contain any studies with human participants or animals, Informed consent: No informed consent is available.